\documentclass{aastex61}

\newcommand\aastex{AAS\TeX}

\usepackage{amssymb}
\usepackage{amsmath}
\usepackage{rotating}
\newcommand{\Lagr}{\mathcal{L}}
\newcommand{\Hist}{\mathcal{H}}

\newcommand{\Ef}{E^{\prime}}

\submitjournal{ApJ}

\shorttitle{\aastex\ sample article}
\shortauthors{Lowell et al.}

\begin{document}

\title{Maximum Likelihood Compton Polarimetry with the Compton Spectrometer and Imager}

\author{A.W. Lowell, S.E Boggs, C.L. Chiu, C.A. Kierans, C. Sleator, J.A. Tomsick, A.C. Zoglauer}
\affil{Space Sciences Laboratory, University of California, Berkeley,  USA}
\author{H.-K. Chang, C.-H. Tseng, C.-Y. Yang}
\affil{Institute of Astronomy, National Tsing Hua University, Taiwan}
\author{P. Jean, P. von Ballmoos}
\affil{IRAP Toulouse, France}
\author{C.-H. Lin}
\affil{Institute of Physics, Academia Sinica, Taiwan}
\author{M. Amman}
\affil{Lawrence Berkeley National Laboratory, USA}

\begin{abstract}

Astrophysical polarization measurements in the soft gamma-ray band are becoming more feasible as detectors with high position and energy resolution are deployed.  Previous work has shown that the minimum detectable polarization (MDP) of an ideal Compton polarimeter can be improved by $\sim 21\%$ when an unbinned, maximum likelihood method is used instead of the standard approach of fitting a sinusoid to a histogram of azimuthal scattering angles.  Here we outline a procedure for implementing this maximum likelihood approach for real, non-ideal polarimeters.  As an example, we use the recent observation of GRB 160530A with the Compton Spectrometer and Imager.  We find that the MDP for this observation is reduced by $20\%$ when the maximum likelihood method is used instead of the standard method.

\end{abstract}

\section{Introduction} \label{sec:intro}

Astrophysical polarization measurements can reveal valuable information relating to the emission mechanism, geometrical configuration, and magnetization of a source, which may not be accessible via spectral, timing, imaging, or photometric analysis.  The primary goal of a polarimeter is to measure the polarization level and angle of a beam.  A secondary goal is to measure the time, energy, angular, or flux dependence of the polarization level and angle.  In the soft gamma-ray band (0.1-10 MeV), where the dominant cross-section for matter-light interactions is Compton scattering, Compton telescopes can deliver strong polarimetric performance owing to the fact that the Compton scattering cross-section is sensitive to the orientation of the photon's electric field vector \citep{lei1997}.

Several space-borne instruments have been used to measure the polarization of astrophysical sources in the Compton regime: INTEGRAL/IBIS and INTEGRAL/SPI observations of the Crab Nebula \citep{ibiscrab,deanspicrab,morancrabchangeibis} and Cyg X-1 \citep{laurentibiscygx1,spicygx1}, the RHESSI observation of GRB 021206 \citep{coburnboggs,rutledgefox,wigger}, and IKAROS/GAP observations of GRB 100826A \citep{yonetoku2011b}, GRB 110301A, and GRB 110721A \citep{yonetoku2012}.  Additionally, the Astrosat/CZTI team has recently reported polarization measurements for 11 bright GRBs \citep{astrosatgrb}.  The standard approach to the data analysis consists of generating a histogram of measured azimuthal scattering angles for qualifying events and fitting a ``modulation curve'' to the data.  While this approach is simple and effective, it disregards information that can be used to further constrain the polarization properties of the beam, such as the Compton (polar) scattering angle, and the initial photon energy.  \citet{kraw} has shown that by combining the Compton scattering angle and photon energy measurements with the azimuthal scattering angle measurement in an unbinned, maximum likelihood analysis, the sensitivity of an ideal polarimeter is improved by $\sim 21\%$ over the standard approach.  In this paper, we describe our efforts to implement this maximum likelihood technique for a real, non-ideal polarimeter, namely the Compton Spectrometer and Imager (COSI) \citep{chiu2015,kierans2017}.  As a test for both the maximum likelihood and standard methods, we use the recent observation of GRB 160530A with COSI, which is described in the accompanying paper ``Polarimetric Analysis of the Long Duration Gamma Ray Burst with the Compton Spectrometer and Imager'', hereafter referred to as P1.

\section{Compton Polarimetry}

The Klein-Nishina equation gives the differential cross-section for Compton Scattering of photons on free electrons at rest:
\begin{equation} \label{eq:kn}
\frac{d\sigma}{d\Omega} = \frac{r_{0}^2}{2}\bigg(\frac{E^{\prime}}{E}\bigg)^2\bigg(\frac{E}{E^{\prime}} + \frac{E^{\prime}}{E} - 2\rm{sin}^2 \theta \rm{cos}^{2} \eta \bigg), 
\end{equation}
where $r_{0}$ is the classical electron radius, $E$ is the initial photon energy, $E^{\prime}$ is the scattered photon energy, $\theta$ is the Compton scattering angle (or polar scattering angle), and $\eta$ is the azimuthal scattering angle defined such that $\eta = 0$ corresponds to scattering along the direction of the initial photon's electric field vector.  After some algebraic manipulation of Equation \ref{eq:kn}, the probability density function (PDF) of scattering with a particular $\eta$ takes the simple form of an offset cosine:

\begin{equation}
p(\eta;E,\theta) = \frac{1}{2\pi} \big[ 1 - \mu(E,\theta)\rm{cos}(2\eta) \big],
\label{eq:azprob}
\end{equation}
where $\mu (E,\theta)$ is the ``modulation'':

\begin{equation}
\mu (E,\theta) = \cfrac{\rm{sin}^2\theta}{\cfrac{E^{\prime}}{E} + \cfrac{E}{E^{\prime}} - \rm{sin}^2\theta}.
\label{eq:mu}
\end{equation}
$E$, $E^{\prime}$, and $\theta$ are all related by the kinematic Compton scattering formula:

\begin{equation}
\Ef = \cfrac{E}{1 + \cfrac{E}{m_e c^2}\big(1 - \rm{cos}\theta\big)},
\label{eq:compton}
\end{equation}
where $m_ec^2 = 511$ keV.  Figure \ref{fig:mod} shows the dependence of $\mu$ on $E$ and $\theta$; Thus, the modulation  is larger at lower energies, and for Compton scattering angles near $\sim 90^{\circ}$.

When a gamma-ray beam is polarized at a level of $\Pi$, where $ 0 \leq \Pi \leq 1$, then a fraction $\Pi$ of the photons from the beam will have their electric field vectors aligned along a specific direction.  The other $1-\Pi$ fraction of the photons will have their electric field vectors randomly oriented.  Thus, for a photon from a beam with polarization level $\Pi$ and polarization angle $\eta_0$, Equation \ref{eq:azprob} becomes:

\begin{equation}
p(\eta;E,\theta,\Pi,\eta_0) = \frac{1}{2\pi} \big[ 1 - \Pi\mu(E,\theta)\rm{cos}(2(\eta - \eta_0)) \big].
\label{eq:idealpdf}
\end{equation}
It is clear from Equation \ref{eq:idealpdf} that photons from a polarized gamma-ray beam will preferentially scatter such that $\eta - \eta_0 = +90^{\circ}$ or $-90^{\circ}$.

\subsection{The Compton Spectrometer and Imager} \label{sec:cosi}

The COSI detector system is comprised of twelve cross-strip germanium detectors, which are capable of measuring interaction locations \citep{lowell2016} and energies \citep{kierans2017} with high accuracy.  For polarization analysis, the relevant measurable quantities are the azimuthal scattering angle, the Compton scattering angle, and the total photon energy.  COSI measures these quantities in the following ways:  The azimuthal scattering angle $\eta$ is determined geometrically, using the positions of the first and second interaction locations along with the known source position (determined by Compton imaging).  The total photon energy $E$ is determined by simply summing all of the measured interaction energies.  Finally, the Compton scattering angle is computed using Equation \ref{eq:compton}, where $\Ef = E - E_1$, and $E_1$ is the energy deposited in the first Compton scatter.

COSI departs from the ideal in several ways.  Most importantly, the detection efficiency is not uniform for all $\eta$, even for specified values of $\theta$ and $E$.  This is mostly a geometric effect, as the detector volume is not azimuthally symmetric about the incident photon beam.  Thus, in directions where more detector material is present, the efficiency will be higher.  Moreover, intervening passive material will have the effect of lowering the efficiency.  Another geometry related effect is the electrode strip segmentation and orientation.  It is currently not possible to assign a depth to events where the same strip was hit more than once, so these events are excluded from the analysis.  Excluding these events reduces the efficiency along directions where a photon would hit the same strip in a detector twice.  Besides geometrical effects, there are other detector effects which can impact the efficiency.  The most important of these is non-uniformity of energy thresholds across channels.  Each strip of each COSI GeD has a threshold which is tuned independently depending on the channel gain and noise.  This can impact the efficiency when photons scatter along directions which lead them to interact in channels with higher or lower than average energy thresholds.

Accounting for these effects through simulation is paramount to the success of polarization analysis. Without careful consideration of these instrumental systematics, spurious signals may arise and real signals may be suppressed.  In our analysis, we use an accurate mass model of the COSI detector system which faithfully represents the distribution of active detector material as well as any passive mass.  Additionally, all simulations are processed by a ``detector effects engine'' \citep{sleator2017}, which adds noise, applies thresholds, performs inverse calibrations, and discretizes the data on a per channel basis.  

\subsection{Standard Analysis Method} \label{sec:sm}

The standard analysis method (SM) aims to fit a generalized form of Equation \ref{eq:idealpdf} to a measured azimuthal scattering angle distribution (ASAD).  The fit function is

\begin{equation}
A - B\textrm{cos}(2(\eta - \eta_0)),
\label{eq:fitfunc}
\end{equation}
where the offset $A$, the amplitude $B$, and the polarization angle $\eta_0$ are free parameters.  The measured modulation $\hat{\mu}$ is
\begin{equation}
\hat{\mu} = \frac{B}{A},
\end{equation}
which can be converted to a polarization level using 
\begin{equation}
\Pi = \frac{\hat{\mu}}{\mu_{100}},
\label{eq:smpi}
\end{equation}
where $\mu_{100}$ is the ``modulation factor'' -- the observed modulation in the case that $\Pi = 1$.  For an ideal polarimeter at a specific photon energy and Compton scattering angle, $\mu_{100}$ is given by Equation \ref{eq:mu}.  However, a real polarimeter typically must accept a range of energies and Compton scattering angles, in which case $\mu_{100}$ represents a weighted average over Equation \ref{eq:mu}.  Moreover, a real polarimeter may not be able to correctly reconstruct all Compton events, thus reducing the value of $\mu_{100}$.  It is usually not possible to determine $\mu_{100}$ experimentally for a given source, due to the difficulty of producing polarized beams with the correct angular distribution and spectral shape in the laboratory.  Therefore, $\mu_{100}$ is estimated via GEANT4 \citep{geant4} Monte Carlo (MC) simulations.

Before the fits take place, there are three pre-processing steps:
\begin{enumerate}
\item Determine the event selections that give the lowest minimum detectable polarization (MDP).  The MDP is given by
\begin{equation}
\textrm{MDP} = \frac{4.29}{\mu_{100}r_s}\sqrt{\frac{r_s + r_b}{t}},
\label{eq:mdp}
\end{equation}
where $r_s$ is the average source count rate, $r_b$ is the average background count rate, $t$ is the observation time, and the factor of 4.29 corresponds to a confidence level of 99\% \citep{weisskopf}.  Note that this equation does not include systematic errors, and so the true MDP must be obtained using MC simulations which have been carefully benchmarked using calibrations.  Nonetheless, Equation \ref{eq:mdp} is a reasonable starting point for event selection optimization.
\item Make a background ASAD and subtract it from the source ASAD.  For transient sources such as GRBs, data from before or after the event may be used.  For persistent sources, the polarimeter must select data from a region of the data space which has similar characteristics, but does not coincide with the source.
\item Correct the background subtracted ASAD for systematics using an unpolarized ASAD.  The unpolarized ASAD is usually obtained by an MC simulation which mimics the source under study with $\Pi_{sim} = 0 $.  The correction is performed by first rescaling each bin of the unpolarized ASAD by the mean value, and then dividing each bin of the background subtracted ASAD by each mean-scaled bin of the unpolarized ASAD.

\end{enumerate}
For an example of this procedure applied to GRB 160530A, please see P1, Section 4.1.

\subsection{Maximum Likelihood Method}

The goal of the maximum likelihood method (MLM) is to find the beam polarization level $\Pi$ and angle $\eta_0$ that maximize the likelihood $\Lagr$
\begin{equation}
\Lagr = \prod_{i=1}^{N}p(\eta_i;E_i,\theta_i,\Pi,\eta_0)
\end{equation}
where $p$ is the conditional probability of measuring the azimuthal scattering angle $\eta_i$ given that we have accurately measured the energy $E_i$ and polar scattering angle $\theta_i$ of event $i$.  For event lists longer than several hundred, $\Lagr$ can easily underflow a double precision floating point number.  To mitigate this problem, the natural logarithm of the likelihood is used
\begin{equation}
\ln \Lagr = \sum_{i=1}^{N}\ln p(\eta_i;E_i,\theta_i,\Pi,\eta_0).
\end{equation}
The values of $\Pi$ and $\eta_0$ that maximize $\ln \Lagr$ also maximize $\Lagr$, since the natural logarithm is a monotonically increasing function.  A hat symbol is used to denote the optimal values, i.e. $\hat{\Pi}$ and $\hat{\eta_0}$.

For an ideal polarimeter, $p(\eta_;E,\theta,\Pi,\eta_0)$ takes the simple form of Equation \ref{eq:idealpdf}.  However, for a real polarimeter, Equation \ref{eq:idealpdf} no longer holds due to the systematic effects of the detector system (Section \ref{sec:cosi}).  The complexity of the MLM thus lies in determining $p(\eta_i;E_i,\theta_i,\Pi,\eta_0)$ for each event $i$ in such a way so as to include the instrument systematics.  Here we outline a simulation based scheme for evaluating $p(\eta_i;E_i,\theta_i,\Pi,\eta_0)$:
\begin{enumerate}
\item Carry out a GEANT4 simulation of the instrument mass model subjected to an \textit{unpolarized} gamma-ray beam with the same coordinates and spectrum as the source under study.  The source coordinates and spectrum used for the simulation must be determined as well as possible, as the response of a polarimeter is often quite sensitive to these parameters.  
\item Define a three-dimensional histogram $\Hist[E,\theta,\eta]$ indexed by energy $E$, polar scattering angle $\theta$, and azimuthal scattering angle $\eta$.  Let the number of $E$, $\theta$, and $\eta$ bins be $j$, $k$, and $l$, respectively.  This histogram will also be referred to as the ``response.''
\item For the $i^{\rm th}$ simulated event, perform the event filtering and reconstruction, determine $E_i$, $\theta_i$, and $\eta_i$, and increment the corresponding cell in $\Hist[E,\theta,\eta]$ by one.
\item The azimuthal scattering angle probability $p(\eta_i;E_i,\theta_i,\Pi,\eta_0)$ for a real event $i$ can now be computed in the following way: take a one-dimensional slice along the $\eta$ axis of $\Hist[E,\theta,\eta]$, and call this slice $g(\eta;E_{j^{\prime}},\theta_{k^{\prime}})$, where $j^{\prime}$ is the index of the $E$ bin containing $E_i$ and $k^{\prime}$ is the index of the $\theta$ bin containing $\theta_i$.  Then the conditional PDF for $\eta$ is
\begin{equation} \label{eq:newpdf}
p(\eta;E_i,\theta_i,\Pi,\eta_0) = \frac{1}{A}\big[g(\eta;E_{j^{\prime}},\theta_{k^{\prime}}) \times \frac{1}{2\pi}\big(1 - \Pi\mu(E_i,\theta_i)\cos(2(\eta - \eta_0))\big)\big],
\end{equation}
where $A$ is a normalization constant chosen so that the area under the total PDF is equal to unity.  Equation \ref{eq:newpdf} can then be evaluated at $\eta_i$ to yield $p(\eta_i;E_i,\theta_i,\Pi,\eta_0)$.
\end{enumerate}
Equation \ref{eq:newpdf} is intuitively simple to understand; the slices $g(\eta;E,\theta)$ encode the effects pertaining to the instrument systematics, and the second term - which is just Equation \ref{eq:idealpdf} - is the ideal PDF.  If this analysis were carried out with an ideal polarimeter, the slices  $g(\eta;E,\theta)$ would be uniform in $\eta$, and Equation \ref{eq:newpdf} would collapse to Equation \ref{eq:idealpdf}.  In essence, the $g(\eta;E,\theta)$ slices represent the acceptance as a function of $E$ and $\theta$.  

An alternative scheme for determining the azimuthal scattering angle probability for each event would be to perform a series of MC simulations with various $\Pi$ and $\eta_0$, and interpolate the responses during the maximization of $\ln \Lagr$ as $\Pi$ and $\eta_0$ are varied.  However, such an approach requires significantly more simulation time in order to achieve adequate statistics in every bin of each response.  Contrast this with the approach outlined above, where only a single simulation of an unpolarized source is needed, and no interpolation based on the values of $\Pi$ and $\eta_0$ is required. In our procedure, interpolation is avoided because the part of the PDF in Equation \ref{eq:newpdf} that depends on $\Pi$ and $\eta_0$ is analytic.

\begin{figure}[t]
\includegraphics[width=\textwidth]{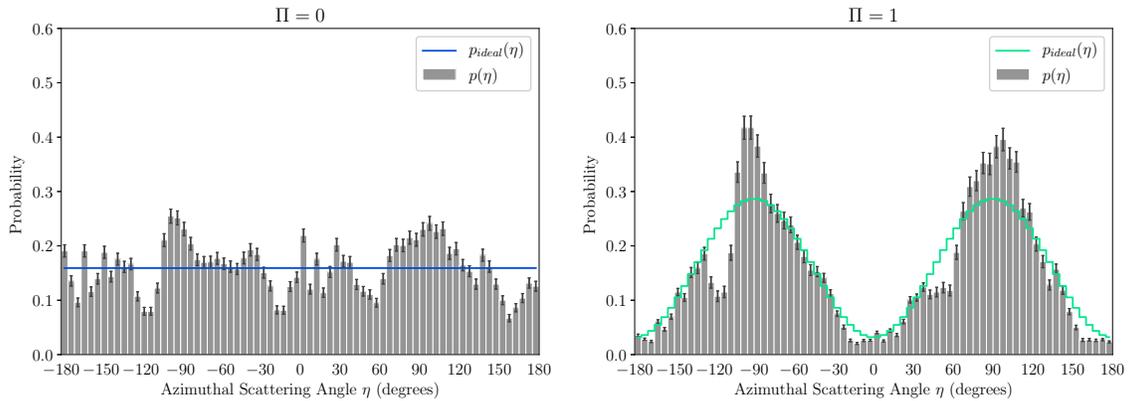}
\caption{The full PDF used for the MLM analysis (gray bars), along with the ideal PDF (blue and green lines) for two cases $\Pi = 0$ (left) and $\Pi = 1$ (right).  The PDFs are drawn for a photon energy of 337.5 keV and a Compton scattering angle of 92.5 degrees.  The slice $g(\eta;E,\theta)$ used for these PDFs is valid over the range $E=325-350$ keV and $\theta = 90-95$ degrees.  The slice $g(\eta;E,\theta)$ used here was taken from the COSI response $\Hist[E,\theta,\eta]$ for GRB 160530A, which occurred $43.5^{\circ}$ off-axis.  Error bars are drawn on the PDFs based on the simulation statistics.} 
\label{fig:pdf}
\end{figure}

Figure \ref{fig:pdf} shows the total PDF for the azimuthal scattering angle for a photon with $E=337.5$ keV and $\theta=92.5$ degrees.  The ideal PDF is overplotted for comparison.  At this energy and Compton scattering angle, the modulation is relatively high.  On the left, where $\Pi = 0$ (unpolarized), the ideal PDF is just a constant, so the full PDF is equivalent to $g(\eta;E,\theta)$.  On the right, where $\Pi = 1$ (fully polarized), the ideal PDF is modulated, and so the full PDF is the normalized product of the modulated, ideal PDF (Equation \ref{eq:newpdf}) with $g(\eta;E,\theta)$.  Clearly, the systematic effects of the detector system (Section \ref{sec:cosi}) distort the PDF from its ideal shape. However, the structure of the ideal PDF still comes through in that where the ideal PDF has peaks, the probability is enhanced, and where the ideal PDF has troughs, the probability is suppressed.  Note that the response used in Figure \ref{fig:pdf} is for the COSI observation of GRB 160530A, which occurred $43.5^{\circ}$ off-axis.

In the presence of background, the probability in Equation \ref{eq:newpdf} must be modified to include a term that represents the background probability distribution:
\begin{equation}
p_{\textrm{total}} = f \cdot p(\eta;E,\theta,\Pi,\eta_0) + (1-f) \cdot p_{\mathrm{bkg}}(\eta;E,\theta),
\end{equation}
where $f = (T-B)/T$ is the signal purity, $T$ is the total number of counts detected, $B$ is the estimated number of background counts in the sample, and $p_{\mathrm{bkg}}$ is the probability of measuring the azimuthal scattering angle $\eta$, given that we have accurately measured the energy $E$ and Compton scattering angle $\theta$, and that the photon originated from a source of background.  A straightforward approach for evaluating $p_{\mathrm{bkg}}$ is to generate a background response $\Hist_{\mathrm{bkg}}[E,\theta,\eta]$ with the same binning as $\Hist[E,\theta,\eta]$, filled with measured background events or simulated background events.  Each $\eta$ slice of $\Hist_{\mathrm{bkg}}[E,\theta,\eta]$ is then normalized so that the bin contents along the $\eta$ axis represent probabilities.  Finally, the background probability for event $i$ can be looked up by retrieving the contents of the bin corresponding to $\eta_{i}$, $E_i$, and $\theta_{i}$.

Once $\hat{\Pi}$ has been found, $\hat{\Pi}$ must be corrected to account for various imperfections of the detector system such as imperfect reconstruction efficiency and measurement error.  Correcting for these effects amounts to determining $\hat{\Pi}$ in the case that $\Pi = 1$, and $B = 0$.  The value of $\hat{\Pi}$ returned by the MLM algorithm under these conditions is referred to as the MLM correction factor, denoted as $\Pi_{100}$.  The measured polarization is then given by:
\begin{equation}
\Pi = \frac{\hat{\Pi}}{\Pi_{100}}.
\label{eq:mlmpi}
\end{equation}
For an ideal polarimeter capable of perfectly reconstructing all events with perfect precision, $\Pi_{100} = 1$.  In reality, some events will be improperly reconstructed and yield a random value for $\eta$, which effectively reduces the measured polarization level.  Additionally, the measurement error on the azimuthal scattering angle will also reduce the measured polarization level.  A plot of $\Pi_{100}$ for the COSI observation of GRB 160530A can be found in P1.

We used the MINUIT minimizer to determine $\hat{\Pi}$ and $\hat{\eta_0}$, and MINOS to determine the errors \citep{minos} for these parameters.  Confidence contours were drawn in the 2D $\Pi$-$\eta_0$ space along paths of constant $2\Delta \log \Lagr$, where $2\Delta \log \Lagr$ is twice the difference between the maximum log likelihood and the log likelihood of a trial point.  This quantity is asymptotically distributed as $\chi^2$ \citep{wilks}, so the confidence level corresponding to a particular value of $2\Delta \log \Lagr$ can be calculated using a $\chi^2$ distribution with two degrees of freedom.

However, the signal purity $f$ has an associated uncertainty stemming from the Poisson distributions underlying $T$ and $B$, which will create additional uncertainty on $\Pi$.  The MINOS errors do not reflect this source of error, because $f$ is held constant during the minimization. Moreover, it is also possible that $\Pi_{100}$ will have a non-negligible uncertainty.  To determine the total uncertainty on the measured, corrected polarization level $\Pi$, we approximated the probability distribution of $\Pi$ by repeatedly simulating the observation.  For each simulated observation, we bootstrap resampled the event list, drew a value of $f$ from it's associated probability distribution, ran the minimizer to determine $\hat{\Pi}$ and $\hat{\eta_0}$, and then divided $\hat{\Pi}$ by a value of $\Pi_{100}$ drawn from its associated probability distribution.  The resulting distribution of $\Pi$ from the simulated observations could then be analyzed numerically to determine confidence intervals or upper limits.

\begin{figure}[t]
\includegraphics[]{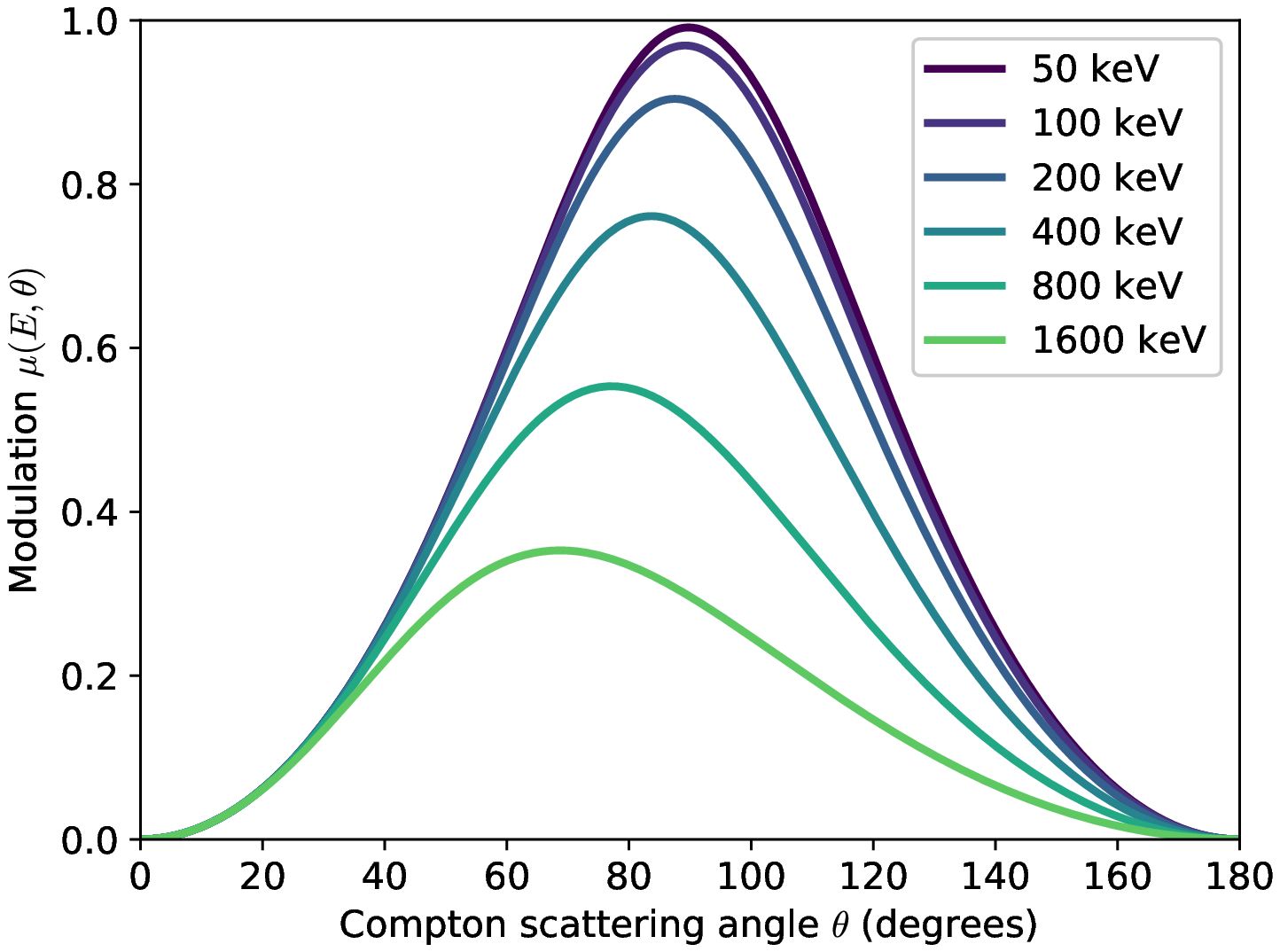}
\caption{The modulation of the azimuthal scattering angle probability as a function of the Compton scattering angle for various photon energies.  At lower energies and for Compton scattering angles near $\sim 90$ degrees, the modulation is strongest.  }
\label{fig:mod}
\end{figure}

The MLM has two main advantages over the SM.  First, more information is used per event.  In the SM, only the azimuthal scattering angle is considered.  In the MLM, the photon energy $E$ and Compton scattering angle $\theta$ are considered as well.  Effectively, each event's contribution to the likelihood statistic is implicitly weighted by Equation~\ref{eq:mu} (Figure \ref{fig:mod}), which is a function of $E$ and $\theta$.  Second, for a realistic observation, the MLM can use more counts in the analysis.  Consider that the first step in the SM is to optimize the statistical MDP (Equation \ref{eq:mdp}).  This amounts to choosing event selections on $E$ and $\theta$ which yield as high a value of $\mu_{100}$ as possible, while still accepting (rejecting) as many source (background) counts as possible\footnote{$E$ and $\theta$ are the most meaningful selections in this context, but other event parameters can and should be optimized as well.}.  In the MLM however, events with any energy or Compton scattering angle can be used.  Therefore, events that were removed during the optimization of the SM analysis can now be included.  Although these events generally represent points on the $\mu$ profile (Equation \ref{eq:mu}, Figure \ref{fig:mod}) corresponding to lower modulation, they are still meaningful contributors to the likelihood statistic.

\begin{figure}[t]
\includegraphics[width=\textwidth]{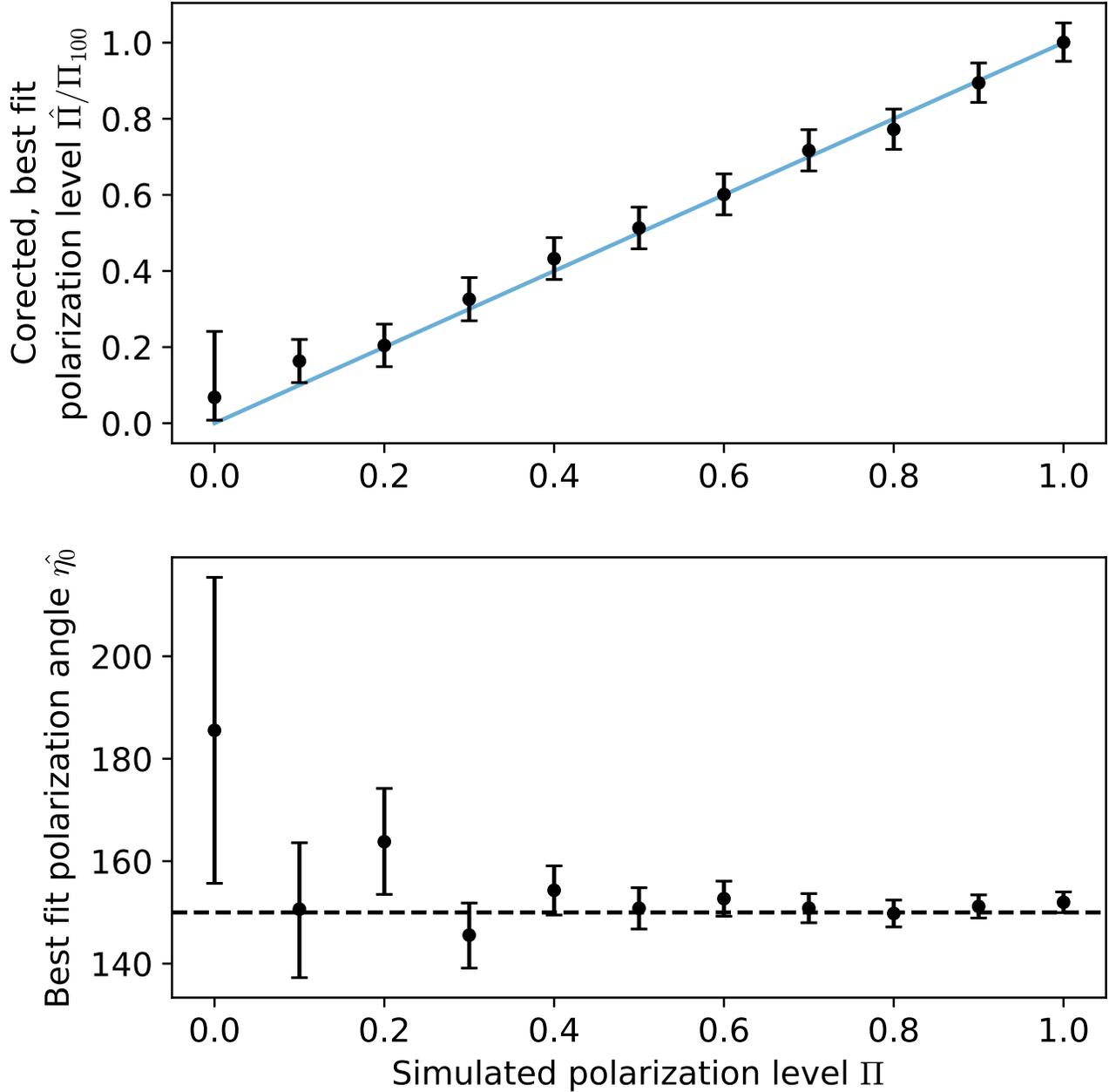}
\caption{The corrected, best fit polarization level $\hat{\Pi}/\Pi_{100}$ (top) and best fit angle $\hat{\eta_0}$ (bottom) as a function of the true, simulated polarization level.  The blue line in the top panel is drawn with a slope of 1 for reference.  These simulations include a background component based on \citet{ling}, with a signal purity of $f=0.72$.  The simulated polarization angle was $\eta_0 = 150^{\circ}$ and is denoted with the dashed line in the bottom plot.  The error bars shown in these plots are $1\sigma$ and were obtained using MINOS.}  
\label{fig:sim}
\end{figure}

We verified the functionality of the MLM by applying the algorithm to simulated data sets, where each simulation had a distinct level of polarization ranging from $\Pi=0$ to $\Pi = 1$.  Figure \ref{fig:sim} shows the corrected, best fit polarization level $\hat{\Pi}/\Pi_{100}$ (top) and $\hat{\eta_0}$ (bottom) for each simulation run.  In the top panel, a blue line with a slope of 1 is drawn for reference.  These simulations used the position and spectrum of GRB 160530A, along with a full sky background based on \citet{ling}.  Events from the background simulation were added to the observation using a signal purity of $f=0.72$, and a background response was generated from the entire background simulation in order to evaluate $p_{\textrm{bkg}}$.
In the top panel, the relationship between $\Pi$ and $\hat{\Pi}$ shows the expected linear behavior of Equation \ref{eq:mlmpi}.  As for the polarization angle, the true value in the simulations was $\eta_0 = 150$ degrees, and is denoted with a dashed line in the bottom panel of Figure \ref{fig:sim}.  As expected, the best fit values for $\eta_0$ become more accurate as the simulated polarization level is increased.  It is clear from Figure \ref{fig:sim} that the algorithm yields reasonable estimates for $\Pi$ and $\eta_0$.

\section{Minimum Detectable Polarization}

The standard measure of a polarimeter's sensitivity is the minimum detectable polarization (MDP), which is the $99^{\rm th}$ percentile of the distribution of $\Pi$ when the true value of $\Pi$ is zero (unpolarized) \citep{lei1997}.  In other words, it is the polarization level above which there is a 1\% chance of measuring a polarization when the source is truly unpolarized.  In the SM, the analytic expression for the MDP is given by Equation \ref{eq:mdp}.  In the MLM, there is currently no known analytical expression for the MDP.  Instead, we calculated the MDP by forming many ($\sim 10,000$), unpolarized trial observations using the same number of source and background counts that were actually observed.  The polarization level of each trial observation was determined using the values of $f$ and $\Pi_{100}$ from our observation of GRB 160530A.
The $99^{\rm th}$ percentile of the distribution of $\Pi$ from the trial observations was then used as the MDP.  Note that because our MC simulations of the unpolarized source use a realistic model of the instrument, the MDP will also include some systematic error.  In order to do a side by side comparison of the MDP between the SM and the MLM, we also used the numerical approach to determine the MDP in the SM, rather than Equation \ref{eq:mdp}.  The procedure was essentially the same as in the MLM case, with the exception that each trial observation was analyzed with the SM, and the value of $\hat{\mu}$ resulting from the fit was divided by $\mu_{100}$ to yield a polarization level (Equation \ref{eq:smpi}).

\begin{figure}[t]
\includegraphics[]{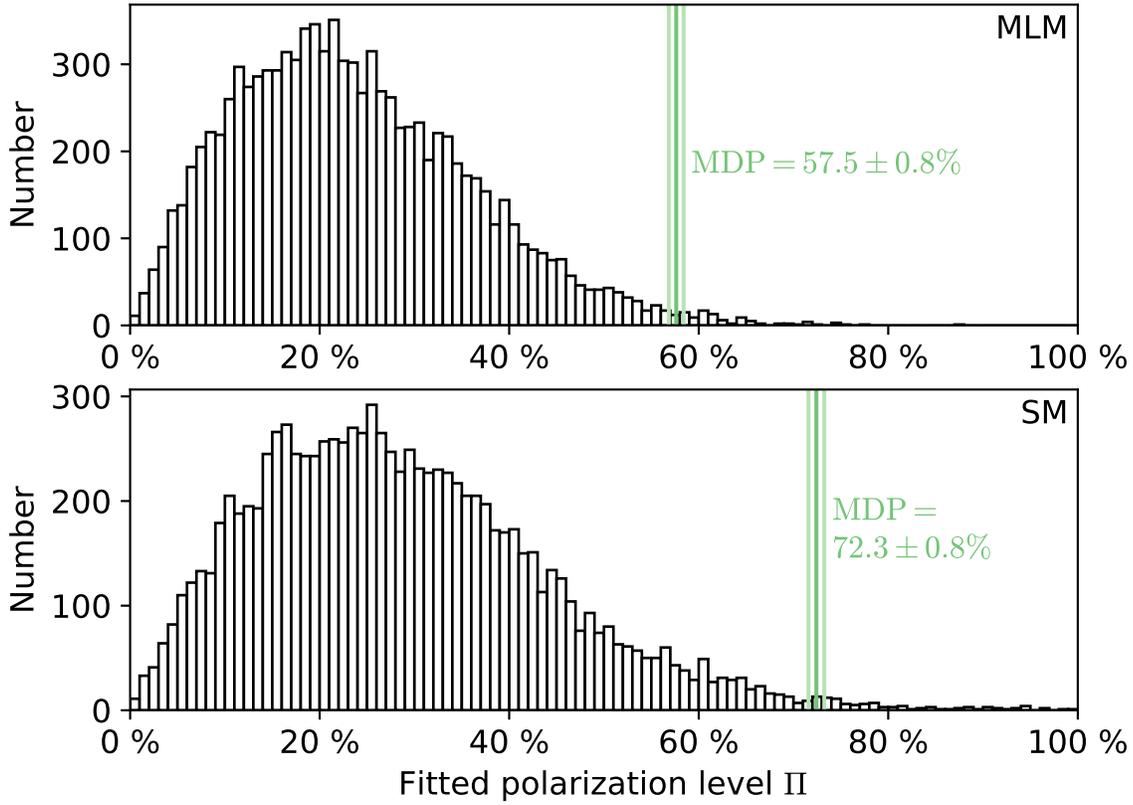}
\caption{Distribution of polarization levels from simulated trial observations of GRB 160530A using the MLM (top) and SM (bottom).  The true polarization level was set to zero for these simulations.  In the MLM case, each trial observation was analyzed in order to determine $\hat{\Pi}$, which was then corrected using the MLM correction factor $\Pi_{100}$ (Equation \ref{eq:mlmpi}).  The resulting polarization level was added to the histogram.  For the SM, the fitted modulation $\hat{\mu}$ was divided by $\mu_{100}$ and then added to the histogram (Equation \ref{eq:smpi}).  The MDP, which is the $99^{\rm th}$ percentile of this distribution, is represented by the dark green vertical lines.  The lighter green lines represent the standard deviation of the MDP, which was obtained by bootstrap resampling the underlying data vector.  The MLM MDP is $20\%$ lower than the SM MDP.}
\label{fig:mdp}
\end{figure}

Figure \ref{fig:mdp} shows the distribution of $\Pi$ using the MLM (top) and SM (bottom) for a simulated polarization level of $\Pi = 0$.  The dark green line denotes the $99^{\rm th}$ percentile of the distribution, and the lighter green lines represent the standard deviation of the $99^{\rm th}$ percentile, obtained by bootstrap resampling of the underlying data vector.  Each trial observation that contributed to the distributions in Figure \ref{fig:mdp} came from realistic simulations of our observation of GRB 160530A.  The number of source counts, number of background counts, and event selections used for the trial observations can be found in Table 2 (SM) and Table 3 (MLM) of P1.  As stated in P1, the MLM was found to improve the MDP by $20\%$ for GRB 160530A.  This improvement is consistent with the findings of \citet{kraw}, where it was shown that application of the MLM can improve the MDP of an idealized Compton polarimeter by $\sim 21\%$.

\section{Discussion}

From the work presented here as well as in P1, it is clear that the MLM is a viable analysis technique that gives a better MDP (sensitivity) than the SM.  Additionally, from the analysis of GRB 160530A in P1, the 90\% confidence upper limit on the polarization level was $\sim 42\%$ lower with the MLM when compared to the SM.  While this can be taken as evidence that the error bars  are more constraining when using the MLM, we have not studied this relationship in detail.

At the outset of this work, our expectation was that the MDP would improve by up to $\sim 21\%$, as \citet{kraw} showed that the MLM improved the MDP by $\sim 21\%$ over the SM with an idealized Compton polarimeter at 100 keV, and under the assumption that the same number of counts was used in both the SM and MLM.  However, as outlined in Section \ref{sec:sm}, the SM event selections must be optimized to give as low an MDP as possible.  After the optimization, some counts will not pass the selections and thus won't be used in the SM analysis.  In the MLM, the event selections for the energy and Compton scattering angle are wide open, so the MLM can use more counts and possibly yield an MDP improvement better than $\sim 21\%$.  If the source spectrum is not monoenergetic, the relative performance between the SM and MLM will also depend on the energy spectrum of the source, and the effective area of the polarimeter at various energies.  This is a consequence of the fact that the modulation of azimuthal scattering angles depends on energy (Equation \ref{eq:mu}, Figure \ref{fig:mod}).  Nonetheless, given the $\sim 21\%$ MDP improvement from \citet{kraw}, as well as our MDP improvement of $20\%$, we conclude that roughly $20\%$ is a good rule of thumb for the maximum improvement one can expect when using the MLM.

The benefits of the MLM come at the cost of higher complexity, as the MLM is more time consuming to implement.  Specifically, generating the polarization response for the MLM requires significantly more simulation time than generating the unpolarized ASAD for the SM due to the much larger number of bins in the MLM polarization response.  However, we believe that the added effort is worthwhile, especially when the number of detected counts is small enough for statistical errors to dominate.  In this work, our polarization response had three dimensions: total photon energy (36 bins), Compton scattering angle (36 bins), and azimuthal scattering angle (36 bins).  Additional dimensions could be used to describe the response, such as the distance between interactions, or position within the detector system.  Including more dimensions would probably improve the performance somewhat, but also substantially increase the number of bins and simulation time needed to fill the bins of the polarization response adequately.

\section{Acknowledgements}
The authors would like to thank Henric Krawczynski for private communications.  COSI is funded by NASA grant NNX14AC81G.  This work is also supported in part by CNES.

\software{GEANT4 \citep{geant4}, MINUIT \citep{minos}, MEGAlib \citep{zogmegalib}, ROOT \citep{root}}

\bibliography{refs}

\end{document}